# Thermal-aware 3D Design for Side-channel Information Leakage

Peng Gu*, Dylan Stow*, Russell Barnes*, Eren Kursun† and Yuan Xie*
*University of California Santa Barbara, CA, USA
{peng_gu, yuanxie}@ece.ucsb.edu
†Columbia University, NY, USA
ek2925@columbia.edu

*Abstract*—Side-channel attacks are important security challenges as they reveal sensitive information about on-chip activities. Among such attacks, the thermal side-channel has been shown to disclose the activities of key functional blocks and even encryption keys. This paper proposes a novel approach to proactively conceal critical activities in the functional layers while minimizing the power dissipation by (i) leveraging inherent characteristics of 3D integration to protect from side-channel attacks and (ii) dynamically generating custom activity patterns to match the activity to be concealed in the functional layers. Experimental analysis shows that 3D technology combined with the proposed run-time algorithm effectively reduces the Sidechannel Vulnerability Factor (SVF) below $0.05$ and the Spatial Thermal Side-channel Factor (STSF) below $0.59$.

## I. Introduction

In recent years the number of attacks aimed at Integrated Circuits (IC) has been increasing rapidly. These attacks include a wide range of strategies spanning invasive, semi-invasive and non-invasive modes, based on the methods that attackers use to acquire valuable data from the ICs [1] [2] [3].

Both invasive and semi-invasive methods involve tampering with the chip to some extent, such as modifying the chip structure or removing the packaging. Such approaches typically require specialized tools and resources. In contrast, noninvasive attacks require fewer resources to perform the attack and conceal traces that the attack ever happened. This poses a greater challenge for hardware security as it enables intrusive monitoring during the normal operation of the chip. As a result, the number of non-invasive attacks has been rapidly increasing in recent years [4].

According to the National Institute of Standards and Technology (NIST) [5], such non-invasive attacks, also known as side-channel attacks, pose a serious threat to the security of a wide range of ICs including cryptographic modules. In cryptographic modules, the cipher hardware is represented as a black box whose internal operations are not observable in theory. However, attackers can bypass the mathematical complexity of the encryption algorithms by extracting and analyzing physical side-channel data such as power dissipation, thermal profiles, electromagnetic radiation, and acoustic traces to determine the secret keys [4].

Recent studies have shown that side-channel attacks can effectively reveal secret key information, as well as substantially reducing the key space that needs to be considered in a bruteforce search [6]. While a large range of countermeasures has been explored [4], none of these techniques can fully prevent side-channel attacks. In most cases, they aim to make the process more difficult to deter attackers.

Thermal Side-channel (TSC) attacks rely on temperature profiling using internal or external sensors to extract critical information from the chip. These attacks have been shown to be effective as a stand-alone side-channel attack mechanism as well as improving the accuracy of other attack types such as Differential Power Analysis [6], which relies on power readings.

The availability of highly sensitive on-chip and off-chip thermal sensors, infrared cameras, and techniques to calculate power consumption from temperature distribution [7] has enhanced the effectiveness of TSC attacks. As a result, sidechannel attacks can be performed by using temperature data without measuring power pins of the chip.

TSC attacks have been reported and analyzed by various researchers in recent years [8], [2], [9], [10], [11] and have been shown to be effective in many different types of ICs ranging from cryptographic units to embedded and general-purpose processors. Built-in thermal sensors have been used to analyze the task scheduling sequence of encryption algorithms [8]. Nefarious programs have been shown to use the TSC as a covert communication channel for transferring sensitive information in FPGAs [9] and multi-core platforms [10]. By decapsulating an embedded microcontroller and attaching a low-cost thermal sensor, recent studies showed that encryption parameters can be acquired from the temperature profile [2].

In recent years, 3D chips have been successfully implemented in a wide range of application areas from embedded chips [12] to memory stacks [13] and general purpose processing units [14]. 3D technology possesses some salient advantages over 2D design in transistor density, interconnect length, heterogeneous integration, and cost reduction. Previous work [15] has proposed a cache design mechanism utilizing wire-length reduction and improved memory bandwidth of 3D technology to make cache-timing side-channel attacks more difficult. To thwart the delay-based side-channel attacks, another work [16] susuccessfully varies the latency of the communication between source and destination ports in 3D Network-On-Chip utilizing dynamic source routing. *However, to the best of our knowledge no previous work has demonstrated implementation of 3D technology for Thermal Side-channel Attack prevention,*

*which utilizes the die-stacking structure of 3D technology.*

This paper proposes a novel solution to protect ICs from side-channel attacks through 3D thermal-aware design, as well as intelligently using dynamic shielding patterns to conceal critical activities on chip. *The proposed 3D integration provides a number of inherent advantages in preventing sidechannel attacks:*

*Invasive Attacks*: Historically, removing packaging to reveal device layers has been an effective step in invasive attacks. However, due to the bonding process, 3D layers can not be removed without harming the normal functionality of the ICs. This protection becomes even more prominent in advanced 3D technologies due to the increased number of thinned device layers [14].

*Semi-Invasive Attacks*: Simple photonic analysis-based attacks [1] require wafer thinning. 3D integration provides protection from such attacks due to the multi-layer stacking that shields the device layers.

*Non-Invasive Attacks*: Device layers and inter-layer bonding materials inherently complicate the radiated side-channel information out of the chip. Thermal profiles of the intermediate layers experience an inherent shielding effect of the layers on top.

In this paper, we propose a novel Thermal-aware Sidechannel Shielding technique, 3D-TASCS, to dynamically camouflage the activity in device layers. This technique relies on an intelligent on-chip controller to track key activity patterns and then generate dynamic shielding patterns to conceal activity. The proposed dynamic shielding algorithm works with hardware thermal management policies by controlling the level of injected noise. Experimental results demonstrate that by leveraging inherent characteristics of 3D integration and employing dynamic shielding patterns, it is feasible to effectively prevent TSC. Also, *the analysis shows that the proposed scheme induces small overhead compared with existing noise injection methods.* The remainder of this paper is organized as follows: Section II discusses the TSC attacks. Section III introduces implementation details of the 3D-TASCS. Section IV specifies the attack model and a metric-based measurement approach for TSC leakage. Section V introduces the algorithms for dynamic thermal pattern generation. Section VI presents experimental analysis of the proposed method. Final conclusions are outlined in Section VII.

## II. THERMAL SIDE-CHANNEL VULNERABILITIES

Thermal Side-channel attacks are typically non-invasive attacks, during which temperature traces are collected to extract key information about the applications running on the chip. Both temporal and spatial thermal data can be correlated with power dissipation for thermal conductivity and capacitance [2].

TSC can be used as a stand-alone attack type as well as in combination with other side-channel attacks. TSC data can be acquired through indirect and direct methods. In the case of indirect utilization, cooling fans have been shown to carry information about the temperature profile of the chip as well as processed data. An approach to extract passwords or RSA keys from this information has been demonstrated [17]. Thermal sensor readings have been used to predict the task execution sequences. This reveals critical information about the activity profiles, such as encryption algorithm scheduling, and it can be applied in combination with other attack types [8].

Recent studies have also shown that it is possible to get the power distribution of a chip by using temperature distribution [7]. This demonstrates that power attack techniques (simple power analysis; differential power analysis) [6] can be carried out via TSC.

In the case of direct utilization, processor core temperatures can be used both as a side-channel and a covert communication channel even when the system implements strong spatial and temporal partitioning [10]. Experiments demonstrate covert thermal channels that achieve up to 12.5 *bps*.

Historically, side-channel attacks have been considered to be more effective on ASICs as they exhibit stronger correlations in the data profiles. However, in recent years, a number of studies have highlighted serious vulnerabilities in general purpose processors as well [17], [8], [10].

A range of side-channel techniques have been demonstrated [2], including high-end thermal sensing equipment with high temporal and spatial resolution, on-chip and off-chip thermal sensors, and infrared camera-based imaging technique. [8] and [10] rely on obtaining temperature information by gaining access to built-in thermal sensors without tampering with the chip. Similarly, trojans are used to gain access to on-chip sensors for thermal side-channel information [3]. Packaging is removed to attach external contact-based thermal sensors directly to the silicon substrate [2]. Infrared cameras acquire power and temperature traces from chips [7]. Charge-CoupledDevice-based thermoreflectance techniques [18] enable high transient ($\approx 500 fps$) and thermal ($10 mK$) resolution imaging of circuits, thus making 2D thermal analysis techniques more powerful. This yields higher precision power and thermal traces than the traditional power pin data used in previous power side-channel attacks.

In this study, we consider all of the aforementioned thermal side-channel attack types: (i) Built-in Thermal Sensor (Noninvasive): Attackers initiate TSC attacks by gaining access to built-in thermal sensors commonly used in ICs. (ii) External Thermal Sensor (Semi-invasive): Attackers acquire temperature profiles by removing packaging and attaching thermal sensors in various locations. (iii) Infrared Thermal Imaging (Semi-invasive or Invasive): Attackers remove packaging and use thermal imaging to analyze temporal and spatial distribution of the temperature data from the chip.

*A. Existing Countermeasures*

In recent years various software techniques were explored to reduce TSC leakage. A secure algorithm for ordering aperiodic tasks with software deadlines has been proposed [8]. While this approach makes the process of inferring the correct task scheduling sequence more difficult, it is not able to fully protect from thermal side-channel attacks as it is still possible to infer task scheduling in certain cases. In [10] a method to restrict access to on-chip thermal sensors was explored, which can eliminate internal sensor reading-based side-channel attacks. However, this technique is not effective in protecting from external sensor or imaging attacks. Furthermore, recent trends in increased accessibility of thermal sensors at the user level limits the overall effectiveness [19]. On the hardware side, noise injection methods have been proposed to protect secret key information. In [20], an FPGA based countermeasure for

Differential Power Analysis was explored. However, this implementation incurs considerable area overhead (44% in the worst case). Similarly, a method [21] exploiting randomized power supply noise injection against Differential Power Analysis has suffered from power overhead challenges (16.5% in the worst case) compared to unprotected
version.

## III. THERMAL-AWARE DESIGN

This study proposes a novel approach that uses 3D technology to protect from TSC by generating dynamic patterns to hide the activities on the device layers. We assume the same 3D manufacturing specs as [14]. As demonstrated in Figure 1, this solution includes: (i) A micro-controller unit that dynamically generates complementary activity patterns to prevent side-channel data leakage. Thermal patterns are generated in a randomized, non-repeating manner such that side-channel attackers cannot extract meaningful information by observing any pattern sequence. (ii) 3D noise generators that run dynamic patterns. On-chip thermal sensors, pattern controllers and noise generators are incorporated along with other functional units using thermal-aware floorplanning.

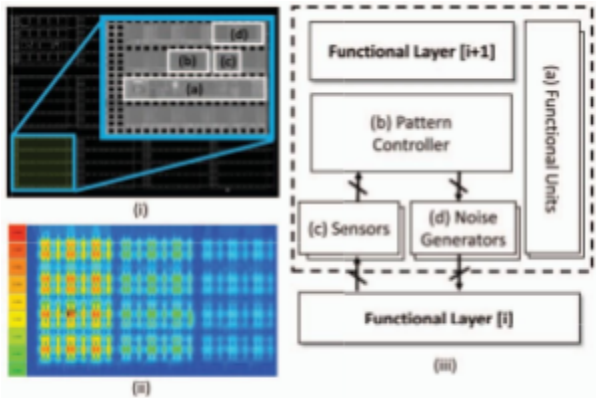

Fig. 1: (i) Layout of a Functional Layer. (ii) Patterns generated by the Pattern Generator Macros. (iii) 3D-TASCS hardware architecture. Note that (a)~(d) are located in the same die.

### A. Thermal-aware Side-channel Shielding Layer Designs

*The proposed technique incorporates targeted strategies to protect from the three attack modes* as discussed in Section II:
*Built-in Sensors*: Instead of associating built-in thermal sensors with individual functional blocks, sensors are placed to read out a composite thermal profile of the device and functional blocks. Therefore, the attackers cannot directly associate temperature readings with specific functional blocks while the overall system thresholds are implemented to avoid
thermal damage or run-away conditions [22].

*External Sensors*: Since external thermal sensors can be placed flexibly at various locations by attackers, the effectiveness of the thermal readings can vary. The proposed algorithm covers all thermal sensor placement options such that noise injected by security layers will decrease the side-channel leakage of any critical areas.

*Infrared Thermal Imaging*: The noise generation in the proposed approach conceals the activity patterns of the functional units from infrared cameras and other imaging devices.

In addition to targeting the individual attack modes separately, the proposed technique works with on-chip power budget and thermal management policies. The power overhead is minimized by intelligently controlling the activity in layers, as will be discussed in Section V.

### B. 3D Design Specification

Three-dimensional integrated circuit technology, as an emerging field, includes several different process integration schemes with varying implementation requirements and integration securities. As a requirement, the protective security layers must be tightly coupled to the functional layer to effectively conceal the thermal information. Thermal or activity information passed between the layers must not be accessible as it could become a source of side-channel leakage. Additionally, an attacker should not be able to remove or disable the protective layers without damaging the functional layer. Several process schemes are compared below:

*System in Package (SiP)*: Although not a true 3D IC implementation, SiP is a readily available packaging technique in which die can be stacked vertically and then connected through wire bonds and package substrate [23]. While SiP is a mature process, the relatively simple manufacturing process compromises the security between layers. In an invasive or semi-invasive attack, the protective layers can be removed or disabled without harming the functional dies. The wire bonds between the die can also be disconnected to power off the security layer, or they can be observed as another source of side-channel information.

*TSV-based 3D IC*: To avoid the vulnerabilities of SiP, a Through-Silicon Via (TSV) 3D process can be employed to tightly integrate the functional and protective layers. Dies are thinned down to 20 $um$ and connected physically and electrically with TSVs [24]. Closer proximity of the security layers increases the difficulty of layer removal and reduces any thermal differential between layers. Information passed on the TSVs is internal to the die stack and not readily available for observation.

*Monolithic 3D*: Although less mature than TSV-based integration, monolithic 3D processes offer the greatest security by effectively integrating multiple semiconductor layers into a single die. Layers are fabricated sequentially on the same wafer, reducing the distance between layers to 100 nm or less. [25]

## IV. MODELING THERMAL SIDE-CHANNEL LEAKAGE

### A. Attack Model

General purpose processors are considered in this paper since they have multiple functional units, so different temporal or spatial instruction execution traces in these functional units will result in different thermal profiles, leading to Thermal Side-channel Leakage. The key assumption is that there are different activity patterns for different functional units, and these different activity patterns can generate thermal patterns that could be distinguished by the attackers. For example, a process carrying critical information $K$ runs on this processor. An adversary could use the Thermal Side-channel attack techniques mentioned in Section II to extract an observation
$F$. Note that the attackers may have run different inputs and have different outputs on the target platform. The attackers may use statistical pattern matching techniques to learn the relationship between the inputs and the outputs. In a real attack, once the attackers have acquired the observation $F$, based on the previous knowledge the critical information *K* may be compromised.

In an encryption co-processor, the thermal traces could be directly collected by the observation of the encryption unit, and the observation could be used to deduce the encryption key. Since there are fewer functional units to track compared to the general purpose processors, the correlations are more vulnerable to being discovered. Observations of the other functional units could be used to assist the discovery of the correlation between the encryption key and thermal traces of the encryption unit.

*B. Metrics*

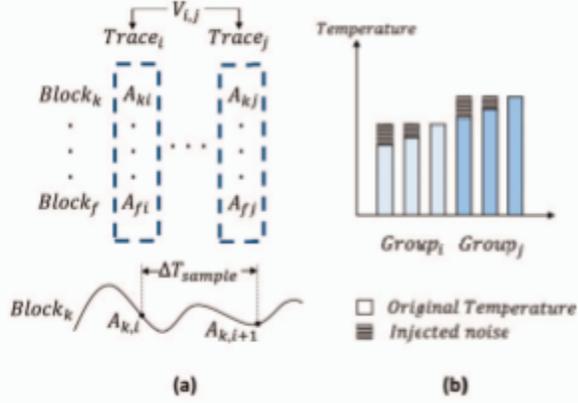

Fig. 2: (a) Execution/Observed traces of different functional blocks used in SVF, and (b) Temperature distribution of different functional blocks used in STSF

Experimental data and models from [14] were used for thermal model verification. In order to quantify the temporal Thermal Side-channel (TSC) Leakage, a statistical metric called Side-channel Vulnerability Factor (SVF) [26] is proposed for analysis. SVF reveals the correlation between the chip's actual execution patterns and the attacker's observations of sidechannel information. In this study, the secret information is defined as the instruction traces (Trace$_{inst}$) of targeted functional blocks in a processor, while the side-channel information is defined as the temperature traces (Trace$_{temp}$) collected by the attackers using techniques mentioned in Section II.

Assuming the attackers collect data periodically for the time interval ($\Delta T_{sample}$), then T race$_{inst}$ and T race$_{temp}$ are two matrices containing instruction count vectors and temperature value vectors of all functional blocks for each $\Delta T_{sample}$, respectively. The execution / observed traces are demonstrated in Figure 2(a).

Temperature traces are usually delayed by $k \cdot \Delta T_{sample}$ from the instruction traces because of the effect of thermal change latency. Since the type of information in these two traces are different, a similar vector is constructed using Equation (1) for each trace, where $Dist(Trace_i, Trace_j)$ represents the Standardized Euclidean Distance between sampled vectors at time interval $i$ and $j$ for the same type of trace. Then, each component in the similar vector $V_{inst}$ is paired with its corresponding component in the similar vector $V_{temp}$ with delayed time $k \cdot \Delta T_{sample}$, and the Pearson Correlation Coefficient is computed based on the list of pairs using Equation (2). Pearson Correlation Coefficient is within the range [−1, +1], where +1 (−1) represents total positive (negative) linear correlation and 0 means no linear correlation. Thus the larger the absolute value of SVF, the more likely the secret information could leak through TSC.

$$V_{i,j} = Dist(Trace_i, Trace_j) \quad i > j, j > 0 \quad (1)$$

$$r = \frac{\sum_{i>j>0}^{n}(V_{inst(i,j)} - \overline{V_{inst}})(V_{temp(i+k,j+k)} - \overline{V_{temp}})}{\sqrt{\sum_{i>j>0}^{n}(V_{inst(i,j)} - \overline{V_{inst}})^2}\sqrt{\sum_{i>j>0}^{n}(V_{temp(i+k,j+k)} - \overline{V_{temp}})^2}} \quad (2)$$

Side-channel Vulnerability Factor (SVF) is a temporal metric for measuring side-channel information leakage. To the best of our knowledge there is no spatial metric for measuring side-channel information leakage. Since thermal distribution of the chips contains 2D spatial information, we propose to use Spatial Thermal Side-channel Factor (STSF) as a complementary measurement to SVF.

For spatial Thermal Side-channel Leakage, the secret information is defined as the relative relationship of the temperature values of functional blocks. Assuming the temperature values of functional blocks are sorted in a sequence: $T_1, T_2 ... T_n$, where $T_i \geq T_j, (i < j)$, to shield the spatial distribution, the sorted sequence is divided into $m$ groups where group $i$ ($i \in [1, m]$) has members: $T1+(i−1) \cdot n/m, T2+(i−1) \cdot n/m ... Ti \cdot n/m$. The temperature values in each group are identical, thus no spatial TSC is leaked within a group as shown in Figure 2(b). Entropy is used to measure the Spatial Thermal Side-channel Factor (STSF) given the number of groups ($m$) as shown by Equation (3), where $-\log(\frac{1}{n!})$ represents the entropy of a sorted sequence, and $-\log\frac{1}{((n/m)!)^m}$ represents the entropy $((n/m)!)^m$ of a sequence which has m groups. The closer the STSF to 0, the less spatial thermal distribution information will leak.

$$r = \frac{-\log(\frac{1}{n!}) - (-\log(\frac{1}{((n/m)!)^m}))}{-\log(\frac{1}{n!})} \quad (3)$$

For thermal analysis, a grid model is used [27] for its flexibility to study thermal patterns by adjusting the granularity of the grid. For a built-in or external thermal sensor, it is assumed that the values are obtained at specific grid points.

V. SHIELDING PATTERN GENERATION

The proposed Thermal-aware Side-channel Shielding technique has design and run-time components to minimize the temporal and spatial temperature variation by injecting activity patterns, thus obscuring the temperature traces. As described in Algorithm 1 and illustrated by Figure 3, the shielding algorithm measures the temperature of a block T$_{sensor}$ from one or more thermal sensors and then estimates the actual block temperature T$_{block}$ (Line1 from Algorithm 1) based on the temperature measurement and the power supplied to the corresponding shielding layers. The block temperature T$_{block}$ is monitored over time to estimate a range [T$_{min}$,T$_{max}$] for the executing instruction stream (Line2). Within this range, a threshold temperature T$_{th}$ is established based on a selected security level S$_{level}$, which is used to balance power with security requirements (Line3 ~ 10). The threshold temperature sets the shielding target; any temperature drop below the threshold will increase the power to the noise generators, thus masking the drop with thermal noise and thereby decorrelating the thermal channel from the instruction trace as shown by Figure 3 (Line11). The power controller can be implemented with a proportional error look-up table for minimal design overhead, or it can employ a full Proportional−Integral−Derivative controller to minimize temperature fluctuations below the threshold temperature. Additionally, the temperature range [$T_{min}$,$T_{max}$] is updated over a slower interval to adjust for long term changes in processor activity (Line2), correcting the threshold temperature and saving power when temperatures decrease.

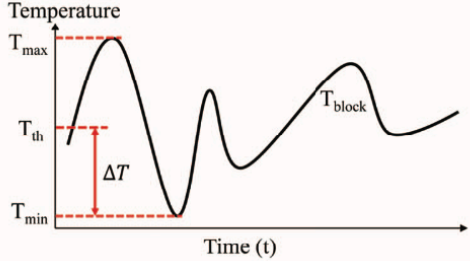

Fig. 3: Illustration of activity patterns of functional units, where $T_{block}$ is the temperature trace of a block in the functional layer.

The algorithm works in conjunction with dynamic voltage and frequency scaling [28] by adjusting the controlled temperature range to the demands of the system thermal management. If overall power or temperature constraints are met, the thermal manager can send control signals to the pattern controller to reduce the shielding output. Activity levels of the pattern generators will be decreased, but functional block temperature will also decrease as the system thermal manager scales the voltage and frequency. The pattern controller will drop the temperature range as the overall system cools, thus shielding the thermal side-channel under thermal and power constraints.

**Algorithm 1: Thermal-aware Side-channel Shielding**

**Input**: $T_{sensor}$, $S_{level}$
**Output**: $P_{generator}$
**Data**: $T_{max}$, $T_{min}$, $T_{th}$

1  Calculate $T_{block}$ from $T_{sensor}$ according to last $T_{th} - T_{min}$;
2  Update $T_{max}$ and $T_{min}$ according to $T_{block}$ for each temperature-adjustment interval;
3  **for** $S'_{level} = S_{level} : -1 : 0$ **do**
4    $\Delta T = T\_table\_lookup(S'_{level})$;
5    **if** $T_{max} - T_{min} < \Delta T$ **then**
6      $T_{th} = T_{max}$;
7    **end**
8    **else**
9      $T_{th} = T_{min} + \Delta T$;
10   **end**
11   $P_{generator} = P\_table\_lookup(T_{th} - T_{min})$;
12   **if** $exceedPowerBudget(P_{generator})$ or $violateThermalThrottling(T_{th} - T_{min})$ **then**
13     Continue;
14   **end**
15   **else**
16     **return** $P_{generator}$
17   **end**
18 **end**

As $T_{th}$ increases, the temporal variation of the temperature trace will be minimized, decorrelating itself from the instruction trace. Also, spatial patterns are effectively hidden if $T_{max}$ is set to be the global maximum value among a group of functional blocks. This camouflages the relative activity of each functional block.

In side-channel secure mode, the pattern generator macros produce noise according to the security level set by the pattern controllers through the power and thermal constraints. This approach works with on-chip thermal management policies, which will activate appropriate actions in the rare cases where the thermal thresholds are exceeded, causing the whole chip to cool down and no TSC information will be leaked.

## V. Experimental Results

In this paper we use a general purpose processing model [29] to demonstrate the effectiveness of 3D-TASCS. It is important to note that this technique can be applied to a wide range of ICs such as embedded systems and ASICs. The benchmarks [30] are simulated using GEM5 [31] where statistics of instruction counts for each functional block are collected every 2ms. The processor is configured as a 4GHz out-of-order CPU, with a 4-way 64KB L1 cache, a 16-way 4MB L2 cache and 2GB main memory. McPAT [32] is used for power analysis, after which the power traces are fed into Hotspot [27] for 3D thermal analysis.

The method described in Section IV-B is implemented along with calculations of Side-channel Vulnerability Factor (SVF) and Spatial Side-channel Leakage Factor as shown in Equation (2) and Equation (3), respectively. For SVF, the proper time delay between instruction traces and temperature traces is first derived to maximize the Pearson Correlation Coefficient by varying the number of delayed time intervals. It is observed that when $k = 13$ according to Equation (2), the maximum correlation point is achieved. Therefore, in the following analysis, temperature traces are delayed for $2.6ms$.

*Mapping temperature increments to power consumption value (P $\_$ table$\_$ lookup in Algorithm 1, Line11):* We use the 3D noise generator macro in the idle mode as the initiation state. Then, we increment the power dissipation of the 3D noise generators and calculate the temperature increments by subtracting the temperature distribution by the corresponding values at initial state.

*Mapping security levels to temperature increments (T $\_$ table$\_$ lookup in Algorithm 1, Line4):* Temperature increments range from $3.5°C$ to $7°C$ with the step size $T_{step} = 0.5°C$, and the change in Side-channel Vulnerability Factor (SVF) values is observed.

Fifteen benchmarks with the highest SVF values ($> 0.1$) are reported in Figure 4. The SVF values of fifteen benchmarks shown in bars are analysed with varying temperature increments from $3.5°C$ to $7°C$ and with maximum temperature increments to shield all thermal side-channel. It can be observed that for the geometric mean ($g\_mean$) using the Thermal-aware Side-channel Shielding algorithm across all benchmarks, SVF values decreases from 0.39 to approximately 0 when temperature increment rises from $3.5°C$ to the maxi-mum temperature increment. However, for all benchmarks the SVF values do not monotonically decrease in each temperature increment. The reason for increase in SVF over the original value is that the injected thermal noise raises the minimum value of temperature trace to the threshold value. This is equivalent to adding low-frequency components and suppressing high-frequency components of temperature traces. When benchmarks exhibit rapid changing instruction traces at peaks but relatively smooth traces at static time (with considerable low frequency components), the noise injected will make the correlation coefficient stronger. After a turning point where the injected low-frequency noise effectively suppresses the high-frequency components, the SVF value decreases when temperature increment rises, as shown in Figure 4. The SVF value decreases as temperature increment rises, as shown in Figure 4. When maximum noise is injected as shown by $max\_avg$, the SVF value drops to 0. In summary, for different benchmarks, the change of SVF values with increasing temperature increment is different. Thus, the temperature increments with SVF values higher than the original SVF value should be eliminated first. Then, the rest of the temperature increments should be sorted according to their corresponding SVF values, and security levels should match the temperature increments through the relative ranking of the SVF values.

Figure 4 also illustrates a power utilization metric of the TASCS algorithm through a scaled SVF. This power utilization metric is calculated as the average power of the pattern generators over the average power of the same generators with the maximum level of noise injection $max\_avg$ as demonstrated in Figure 5. The metric of power utilization is used to scale each SVF value by taking the product of these two values. The products are represented by the dots for each benchmark in Figure 4 and reflects the trade-off between security and power utilization. It can be observed that SVF values with low temperature increments ($T \leq 4.5°C$) are scaled lower than SVF values with high temperature increments ($T \geq 6.5°C$). However, the distribution of scaled SVF values (local minimum values 2.4 and 0.1 with $3.5°C$ and $7.0°C$ temperature increments respectively) is the same as the original distribution of SVF values for all benchmarks. This implies that the SVF values generated by the proposed algorithm are in compliance with power utilization scaling. Furthermore, it indicates that in security modes, low temperature increments ($T \leq 4.5°C$) and high temperature increments ($T \geq 6.5°C$) could be chosen to improve power utilization.

Based on the above observations, optimization could be incorporated into run-time management with consideration of power efficiency. Since the increased temperature will not exceed the maximum temperature of the original functional layer's temperature trace, the thermal constraint is satisfied. The power budget of Algorithm 1 is analyzed depending on the trend of SVF when temperature increment increases. By definition, the lower the SVF value, the less linear correlation could exist between instruction traces and temperature traces. Thus, higher security levels should relate to lower SVF values.
(i) When SVF values decrease monotonically with rising temperature increment, there is a direct trade-off between activity level and security. Therefore, power consumption could be traded for higher TSC protection. (ii) When SVF values increase monotonically with rising temperature increment, a higher security level is associated with lower power consumption. In this situation, the TASCS algorithm simply uses the lowest thermal noise injected to reduce SVF values.
(iii) For benchmarks with irregular changes of SVF values when temperature increment increases, higher security level is associated with temperature increment of lower SVF values. In this situation, decreasing the security level does not necessarily reduce the power consumption.

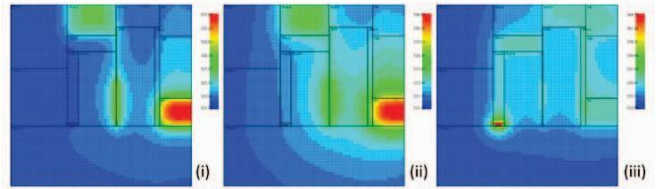

Fig. 6: The thermal profile of benchmark $gcc$ on the functional units with and without the proposed algorithm: (i) Thermal distribution without 3D-TASCS. (ii) Thermal distribution with 3D-TASCS functionality turning off. (iii) Thermal distribution with 3D-TASCS turning on.

Figure 6 shows the thermal profile with and without TASCS algorithm. By increasing the number of groups $m$ =(1, 2, 4, 8), Spatial Thermal Side-channel Factor (STSF) as defined in Equation (3) can be reduced from 0.82 to approximately 0. Lower STSF means higher security level, since the relative relationship among the temperature values of different functional blocks are obscured. The STSF could be reduced to 0.59 on average across different benchmarks using the shielding algorithm.

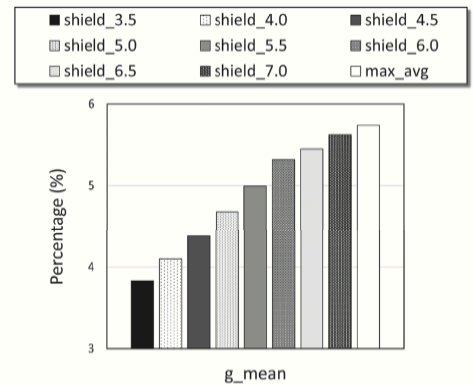

Fig. 7: Metric of Power Overhead for shielding temporal side-channel leakage.

Figure 7 shows the power overhead associated with different shielding factors. The metric for power overhead is calculated as the average power of pattern generations over the total system power. For the same shielding level, $g\_mean$ of the metric is 3.83% for $shield\_3.5$ and upper-bounded by 5.74% for $max\_avg$. This demonstrates that TASCS algorithm can provide effective side-channel shielding with minimum power overhead.

The hardware overhead of the proposed method is small since the design takes use of the already existing 3D design instead of adding a new layer. The noise generators are assumed to be simple buffer chains consisting of back to back inverters that continuously oscillate. They are sparsely distributed among the critical areas to be protected. Compared to the previous randomized method for noise injection [20], this extra area overhead is trivial.

## VI. Conclusion

This paper proposes a 3D Thermal-aware Side-channel Shielding (TASCS) technique to proactively conceal critical activities in the functional units. This technique leverages inherent characteristics of 3D integration and dynamically generates custom activity patterns to shield from thermal side-channel attacks. Experimental analysis shows that this approach reduces the Side-channel Vulnerability Factor (SVF) below 0.05 and the Spatial Thermal Side-channel Factor (STSF) below 0.59. Furthermore, the proposed approach has minimum power overhead, which is less than 5.7% in the worst case.

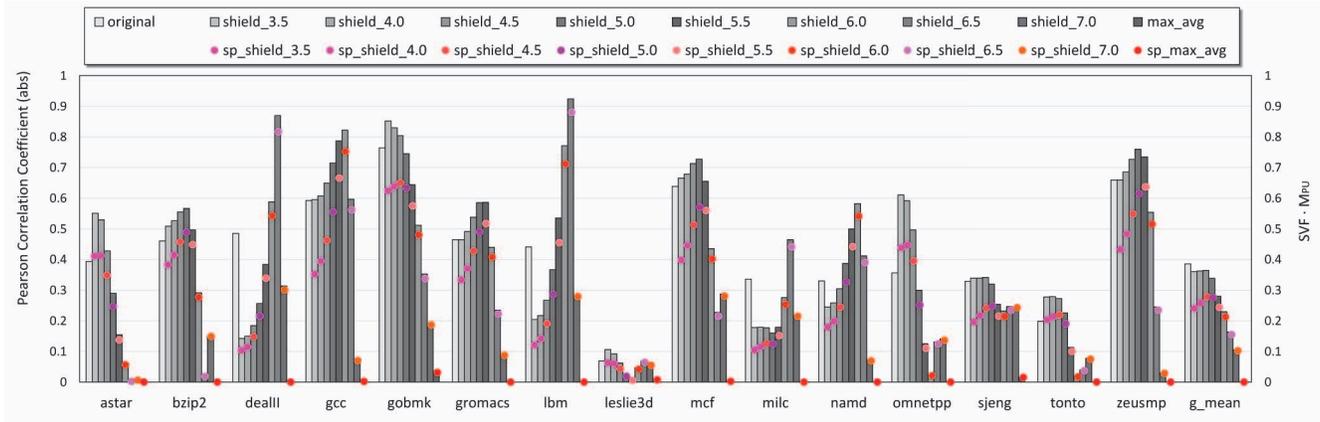

Fig. 4: Side-channel Vulnerability Factor (SVF) as shown in the bars, which computes the Pearson Correlation Coefficients between the similar vectors of the side-channel information (the temperature traces of functional blocks) and the oracle data (the number of instructions executed by the corresponding functional blocks) according to Equation (2). Temperature increment ($T_{th} - T_{min}$) in Algorithm 1 is varied from $3.5°C$ to $7.0°C$ with the step size $0.5°C$. $shield\_n$ represents different levels of noise injection and $n$ represents temperature increment. $max\_avg$ method injects thermal noise to let the average value of the temperature trace equal to its maximum value, so that its SVF for each benchmark is approximately 0 as shown in the figure. The product of SVF and metric of power utilization ($sp\_shield\_n$) is shown in dots, which reflects the trade-off between security and power-efficiency.

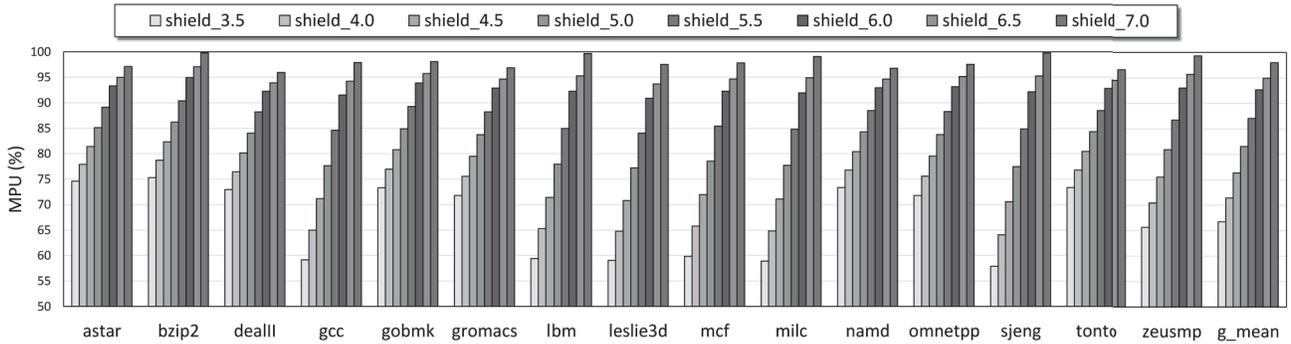

Fig. 5: Metric of Power Utilization (MPU) for shielding temporal side-channel leakage.